# Thermal and Electrical Properties of (Cr,Mo,Ta,V,W)C High-Entropy Carbide Ceramics

Ali Sarikhani[1,*], Steven M. Smith[2], Suzana Filipovic[1], William G. Fahrenholtz[2,*], Gregory E. Hilmas[2]

[1]Missouri University of Science and Technology, Materials Research Center, Rolla, MO, USA

[2]Missouri University of Science and Technology, Materials Science and Engineering, Rolla, MO, USA

**Abstract**

The synthesis and characterization along with the resulting properties of fully dense (Cr,Mo,Ta,V,W)C high entropy carbide ceramics were studied. The ceramics were synthesized from metal oxide and carbon powders by carbothermal reduction, followed by spark plasma sintering at various temperatures for densification. Increasing the densification temperature resulted in grain growth and an increase in the lattice parameter. Thermal diffusivity increased linearly with testing temperature, resulting in thermal conductivity values ranging from ~7 W/m•K at room temperature to ~12 W/m•K at 200 °C. Measured heat capacity values matched theoretical estimates made using the Neumann-Kopp rule. Room temperature electrical resistivity decreased from ~137 to ~120 μΩ•cm as the excess carbon decreased from 5.4 to 0.1 vol%, suggesting an enhanced electronic contribution to thermal conductivity as excess carbon decreased. All specimens exhibited a Vickers hardness of ~29 GPa under a 0.49 N load. These results underscore the tunability of this high entropy carbide system.

* Corresponding authors: A. Sarikhani (as5kw@mst.edu), W. G. Fahrenholtz (billf@mst.edu)

**Introduction**

High-entropy carbides (HECs) are an emerging class of ceramics that combine multiple metal carbide components into a single-phase solid solution [1-3]. These materials exhibit exceptional thermal and mechanical properties, including high melting points (>3500 °C), Vickers hardness values exceeding 25-30 GPa, chemical stability under extreme conditions, and can, in some cases, be stabilized by configurational entropy [2,4,5]. Inspired by the development of high-entropy alloys [6,7] and high-entropy borides [8], HECs have gained attention for applications in thermal protection components in hypersonic systems and plasma-facing materials in fusion energy systems [9-11].

Transition-metal carbides (TMC) possess a wide range of non-stoichiometry on the carbon sublattice and can also accommodate oxygen on those sites; accordingly, the carbon sublattice occupancy is denoted as $TMC_xO_y$. Studies on individual monocarbides show that carbon vacancies can order, while oxygen can dissolve to form oxycarbide solid solutions that retain the rock salt lattice; for example, carbothermal synthesis in the Hf-C-O system stabilizes $HfC_xO_y$ compositions spanning roughly $HfC_{0.98}O_{0.02}$ to $HfC_{0.89}O_{0.11}$ in the range of 1650-1750 °C; while analogous $ZrC_xO_y$ phases also form in the Zr-C-O system [12-14]. The phase-equilibria and site-occupancy behaviors in the binaries provide a framework for interpreting how carbon deficiency and oxygen incorporation may influence lattice parameters, defect scattering, and transport in multicomponent HECs, even though direct phase-equilibria maps for HECs are not yet available [15].

Among the various HEC compositions, (Cr,Mo,Ta,V,W)C has been shown to form a single-phase rock salt crystal structure with a homogeneous distribution of different species on the metal sites, similar to several previously studied HECs [16-18]. These ceramics have demonstrated thermal conductivities in the range of ~7-12 W/m•K at temperatures below 300 °C and electrical resistivity of ~120-140 μΩ•cm at room temperature [19-21]. Vickers hardness values of ~25-32 GPa have been reported depending on composition, densification conditions, and the resulting grain size [6,18]. Prior studies have established that carbothermal reduction followed by spark plasma sintering (SPS) from 1600-2000 °C is an effective route to synthesize dense HEC ceramics, but challenges remain in controlling carbon content and phase purity [5,17,20,22].

While much of the existing work has focused on phase formation and mechanical performance, fewer studies have systematically investigated how sintering temperature and carbon additions influence densification, lattice parameters, and thermal/electrical transport in HECs [4,21,23]. In particular, excess carbon can segregate as a second phase, increase electrical resistivity, and reduce thermal conductivity by enhancing phonon and electron scattering, while not significantly changing the hardness [21,24]. Previous studies show that amorphous or thin graphitic films at grain boundaries increase electron and phonon scattering, raising electrical resistivity and lowering thermal conductivity via interfacial resistance [25,26].

This work provides a quantitative structure-property-processing relationship for HEC ceramics and clarifies how carbon additions, along with densification temperature impact microstructure and properties.

**Methods and Materials**

Commercial chromium oxide ($Cr_2O_3$, 99.5%, 0.7μm; Elementis, Corpus Christi, TX), molybdenum oxide ($MoO_3$, 99.9%, 6 μm, US Research Nanomaterials, Houston, TX), tantalum oxide (Ta$_2$O$_5$, 99.8%, 1 - 5 mm, Atlantic Equipment Engineers, Upper Saddle River, NJ), vanadium oxide (V$_2$O$_5$, 99.6%, −10 mesh, Alfa Aesar), tungsten oxide (WO$_3$, 99.9%, ~80 nm, Inframat Advanced Materials, Manchester, CT), and carbon black (C 120, Cabot Alpharetta, GA) were used as starting powders. Starting powders were batched based on optimization of the carbon to oxide ratio specific to this composition. Carbon-deficient compositions were prepared by reducing the carbon to metal oxide ratio based on generalized carbothermal reduction reactions for the different transition-metal oxides (Reactions 1-3).

$$TMO_3 + (x - y + 3)\, C \rightarrow TMC_xO_y + (3 - y)\, CO \qquad (1)$$

$$TM_2O_3 + (2x - 2y + 3)\, C \rightarrow 2\, TMC_xO_y + (3 - 2y)\, CO \qquad (2)$$

$$TM_2O_5 + (2x - 2y + 5)\, C \rightarrow 2\, TMC_xO_y + (5 - 2y)\, CO \qquad (3)$$

The generalized reactions can be combined into the overall descriptive process described by Reaction 4.

$$0.5\, Cr_2O_3 + MoO_3 + 0.5\, Ta_2O_5 + 0.5\, V_2O_5 + WO_3 + (5x - 5y + 12.5)\, C \rightarrow$$
$$5\, (Cr_{0.2}, Mo_{0.2}, Ta_{0.2}, V_{0.2}, W_{0.2})C_xO_y + (12.5 - 5y)\, CO \qquad (4)$$

This notional reaction assumes that decreasing the carbon addition enables oxygen dissolution into the HEC lattice. The ratio of carbon to total transition metal content (five metals in equimolar amounts) was assumed to be 1:1 for the optimized reaction. Finally, the notional reactions assumed that sum of the lattice sites occupied by carbon ($x$) and oxygen ($y$) were assumed to be equal 1 (i.e., $x + y = 1$) so that all of the carbon lattice sites were filled by either carbon or oxygen. Three notional compositions were produced. The first assumed that the stoichiometric amount of carbon was incorporated, which would result in $x=1$ and $y=0$. The second composition was deficient in carbon by 2.5 wt% that would result in *x=0.96* and *y=0.04*. The third composition as deficient in carbon by 7.5 wt%, which results in *x=0.87* and *y=0.13*. If the batching only resulted in carbon vacancies (i.e., no oxygen dissolved into the carbon vacancies in the lattice, then the nominal compositions would be $(Cr,Mo,Ta,V,W)C_{1-\delta}$ where $x=1-\delta$. In the actual HECs, carbon vacancies are likely to be only partially occupied by oxygen, which is somewhere between the extremes represented by $(Cr,Mo,Ta,V,W)C_xO_y$ and $(Cr,Mo,Ta,V,W)C_{1-\delta}$.

The powders were mixed by high energy ball milling (SPEX, 8000D MIXER/MILL, Glen Mills Inc.) in a WC jar using WC media. The mass ratio of WC media to powder was 15:2. Two 20-gram batches of each composition were pressed into pellets under an ~100 kg load using 2.5 cm diameter steel dies after 60-mesh sieving. The pellets were reacted at 1610 °C for 3 hours under a vacuum pressure of ~ 13.3 Pa. Finally, the reacted powder was passed through a 100-mesh sieve and then split in two parts for densification by a SPS process, as described previously [27]. The SPS process began under vacuum (<6.0 Pa), ramping to 1600 °C at 100 °C/min under a uniaxial pressure of 15 MPa. After a 5-min dwell, uniaxial pressure was increased to 50 MPa over the time span of 1 min, followed by heating to the final densification temperatures of either 1700 or 1950 °C at 100 °C/min. The dwell at the peak temperature was 10 min. During cooling, pressure was reduced to 25 MPa over one minute while the temperature was decreased to 1200 °C at 50 °C/min. Ram displacement was recorded to extract densification and thermal expansion data.

X-ray diffraction (XRD; X-Pert MPD, Philips) was performed at room temperature for the sintered pellets. Analysis (X'Pert High Score, Malvern Panalytical) and Rietveld refinement (RIQAS, Materials Data, Inc.) software were used to index patterns and determine lattice parameters. Theoretical densities were calculated based on the nominal equiatomic metal ratios from the batched compositions and lattice parameters determined by Rietveld refinement of the XRD patterns of the densified ceramics. Relative densities were estimated from these theoretical densities and measured bulk densities. Bulk density was measured using the Archimedes method according to ASTM C20 with water as the immersing medium.

Thermal diffusivity was measured for specimens that were 12-mm squares with thicknesses between 0.8 mm and 2.1 mm. Specimens were coated with graphite spray to promote uniform energy absorption during testing. Thermal diffusivity was measured by laser flash (Flashline L-S2, Anter Corporation) with measurements at nominally 25 °C intervals between room temperature and 200 °C (ASTM E1461). Thermal conductivity was calculated by combining the measured diffusivity values with bulk density and heat capacities as described by Equation 5:

$$\kappa = DC_p\rho_b \text{ [W/m•K]} \quad (5)$$

where κ is the thermal conductivity (W/m•K), D is thermal diffusivity (cm$^2$/s), Cp is heat capacity (J/kg•K), and $\rho_b$ is bulk density. Bulk densities were assumed to be constant in this temperature range. The heat capacity was approximated using the Neumann-Kopp rule, which is expressed as:

$$C_{pm}(Cr_{0.2}, Mo_{0.2}, Ta_{0.2}, V_{0.2}, W_{0.2})C_{(s)} =$$
$$0.2\ C_{pm}CrC_{(s)} + 0.2\ C_{pm}MoC_{(s)} + 0.2\ C_{pm}TaC_{(s)} + 0.2\ C_{pm}VC_{(s)} + 0.2\ C_{pm}WC_{(s)} \quad (6)$$

where the heat capacities at different temperatures for the thermodynamically stable monocarbide, TaC, were extracted from the NIST database [28]. Heat capacities for the unstable monocarbides (CrC, MoC, VC, and WC with rock salt structures) are not reported in this database. Instead, the Neumann-Kopp rule was applied again using the molar heat capacities of the constituent metal and carbon.

The electrical resistivities were measured using the Van der Pauw method (ASTM F76). Each specimen was clamped using four stainless steel screws on all four sides, as described in the reference [24]. A current was applied across two leads while the other two leads were used for measuring the emergent voltage. A switch box was used to rotate the electrode configuration among all permutations, which resulted in resistivity measurements that were independent of length and width of the specimen. The resistivity was calculated using Equation 7:

$$R_{||} \cdot R_{\perp} = \left(\frac{\rho_e}{t}\right)^2 \quad [\mu\Omega^2] \tag{7}$$

where $R_{||}$ and $R_{\perp}$ are resistances obtained from two perpendicular measurements, $\rho_e$ is the electrical resistivity (μΩ•cm), and t is the thickness which must be uniform for the Van der Pauw method. The measurements were repeated for three different applied currents of 0.2 A, 0.3 A, and 0.4 A.

The Wiedemann-Franz law was used to estimate the electron contribution to thermal conductivity. The law assumes that the electron contribution to thermal conductivity ($\kappa_{electron}$) is proportional to electrical conductivity ($\sigma$) and temperature ($T$) as shown in Equation 8:

$$\kappa_{electron} = \sigma LT = \frac{LT}{\rho_e} \quad [W/(m.K)] \tag{8}$$

where $L_o$ is the theoretical Lorenz number approximated as $2.44 \times 10^{-8} \; W.\Omega/K^2$ at room temperature and $\sigma$ is equal to the inverse of electrical resistivity ($\rho_e$). Total thermal conductivity ($\kappa$) was assumed to be the sum of the electron and phonon ($\kappa_{phonon}$) contributions.

The hardness was determined by Vickers indentation (Duramin 5, Struers, Cleveland, OH) following ASTM C1327 with loads ranging from 0.49 N (0.05 kg) to 9.8 N (1.0 kg) and a dwell time of 15 seconds. Indentations were made on polished cross-sections. For each load, at least five valid indentations (with no major cracks or spalled edges) were measured using an optical light microscope (KH-3000, Hirox-USA, Hackensack, NJ). Then, the hardness was determined using the relation,

$$HV_x = 1.8544 \frac{F}{d^2} \ [GPa] \tag{9}$$

in which, $HV_x$ is the hardness at a load of x kg, F is the indentation load, and the d is the average length of the two diagonals of the indent.

Table I represents the descriptions of the specimens used for optimizing this HEC by varying the batch carbon addition with a sub-stoichiometric C ratios of 2.5 wt% and 7.5 wt%, each sintered at SPS temperatures of 1700 and 1900 ºC. This table also includes the bulk and theoretical densities, calculated from the refined lattice parameters and equiatomic EDS percentages for the five transition metals. The last column is the average of excess C estimations from multiple SEM images for each specimen using image analysis software (ImageJ, National Institutes of Health).

**Table I.** Specimen descriptions, batch carbon addition, densities, and micrographs resultant excess carbon of (Cr,Mo,Ta,V,W)C high entropy carbide ceramics.

| Specimen | Sintering Temperature (°C) | Batch C (wt%) | Theoretical Density (g/cm³) | Bulk Density (g/cm³) | Relative Density (%) | Microstructure Avg. Excess C (vol%) |
|---|---|---|---|---|---|---|
| 1 | 1700 | -2.5 | 10.14 | 10.04 | 99 | 5.4 |
| 2 | 1950 | -2.5 | 10.42 | 10.38 | 100 | 4.9 |
| 3 | 1700 | -7.5 | 10.16 | 10.30 | 100±1 | 1.8 |
| 4 | 1950 | -7.5 | 10.51 | 10.61 | 100±1 | 0.1 |

**Results and Discussion**

*Microstructure and Composition*

Room-temperature XRD confirmed that the carbides had the rock salt structure with only trace amounts of impurity phases in some specimens. Room-temperature XRD patterns for all four (Cr,Mo,Ta,V,W)C specimens index to the rock salt FCC structure, showing the expected all-odd or all-even (hkl) peaks and no crystalline impurity peaks in all three specimens (Fig. 1). In specimen 2, a weak peak at $2\theta \approx 26°$ was observed and was consistent with the (002) peak of graphitic carbon, while the diffuse background elsewhere may be due to the presence of amorphous carbon; the graphitic signal likely reflects processing conditions in which the combination of temperature and excess C permitted partial crystallization of the carbon. Overall, the diffraction patterns indicated formation of a predominantly single-phase rock salt carbide phase across the processing conditions, with only specimen 2 exhibiting a trace graphitic-carbon impurity at the XRD detection limit.

Rietveld refinement showed a monotonic increase in the cubic lattice parameter with decreasing microstructural excess carbon, consistent with partial filling of carbon vacancies and increased metal–carbon repulsion. Quantitatively, the refined parameter increased from ~4.402 Å for specimen 1 to ~4.409 Å for specimen 4 (Fig. 1, inset). The horizontal axis in the inset represents the average of excess C estimations from multiple SEM images for each specimen using image analysis software. The uncertainties were taken from the refinement fits. Thus, structural refinement links minimization of excess carbon in the microstructure directly to a subtle, but systematic, lattice expansion.

      Theoretical densities were computed from refined lattice parameters assuming equiatomic metal ratios and full occupancy of carbon lattice sites. The bulk densities were measured by the Archimedes method (ASTM C20), yielding relative densities of ~100% for all four specimens (Table 1). Mass-loss balance during carbothermal reduction aligned with reaction stoichiometry. The calculated mass losses due to CO formation were 7.06 g for the 2.5 wt% C-deficient batch and 6.81 g for the 7.5 wt% C-deficient batch, while measured average losses were 7.72 g and 7.31 g, respectively. The small positive offsets in actual mass loss compared to calculated loss were attributed to handling/sieving and powder loss to vacuum during reaction. The bulk density measurements combined with the theoretical densities estimated from lattice parameters confirmed that porosity was negligible and that subsequent property trends were due to differences in carbon content that were not confounded by incomplete densification.

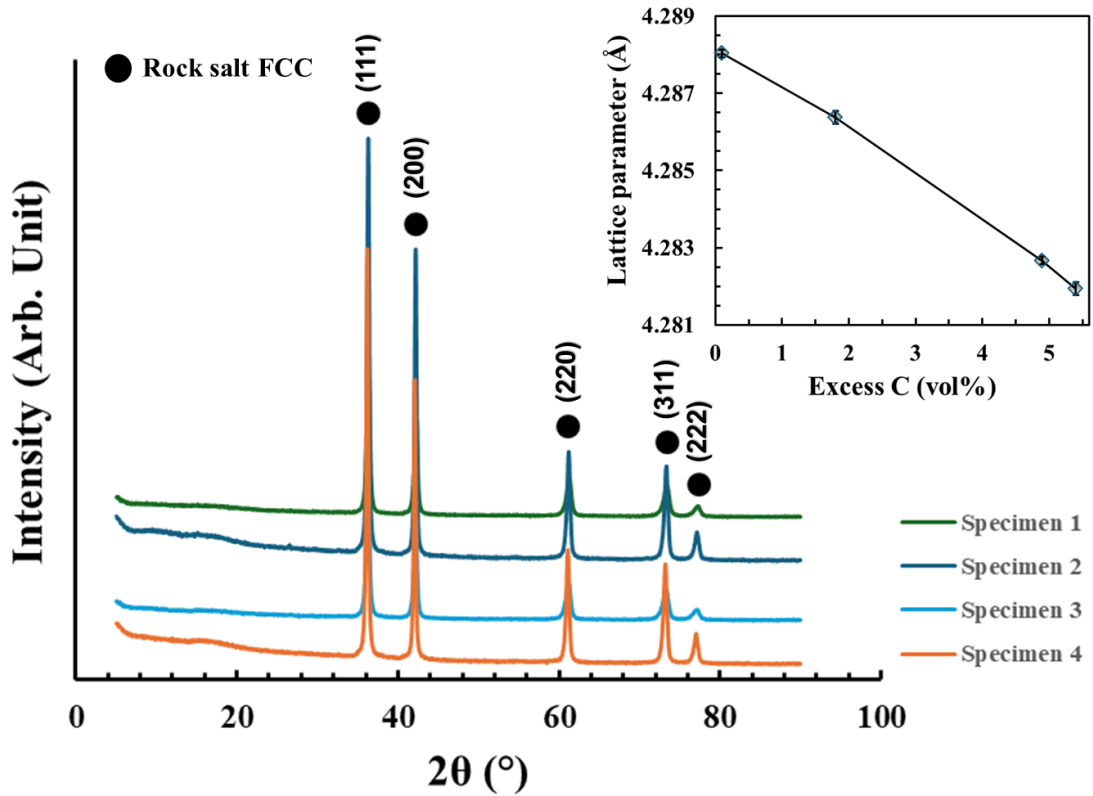

**Figure 1.** XRD peak patterns of the four specimens of (Cr,Mo,Ta,V,W)C with 2.5 and 7.5 wt% sub-stoichiometric batch carbon, sintered at two SPS temperatures of 1700 and 1950 ºC, along with the refined lattice parameters in the inset.

Back-scatter SEM of specimen 2 showed an equiaxed, pore-free microstructure with only trace dark contrast from amorphous carbon (~ 0.1 vol%) and with visible orientation contrast from electron channeling (Fig. 2d). Quantitatively, this specimen achieved ~100% relative density (Table 1), and its higher SPS temperature promoted grain growth and improved interparticle bonding compared with lower-temperature counterparts. Based on microstructure, specimen 4 served as the reference because it represented an optimized combination of relative density and minimal excess carbon content.

The excess C contents in the microstructures were 5.4 vol% for specimen 1 (1700 °C, 2.5 wt% C), 4.9 vol% for specimen 2 (1950 °C, 2.5 wt% C), and 1.8 vol% for specimen 3 (1700 °C, 7.5 wt% C), compared to 0.1 vol% in the optimized specimen, i.e., specimen 4 (Fig. 2a-d). Some of the differences in carbon content in the final ceramics could be due to the higher densification temperatures that promoted carbothermal reduction of metal oxides by increasing the thermodynamic driving force, allowing more carbon to be consumed in carbide formation and to dissolve into carbon vacancies in the lattice through faster transport kinetics at the higher temperature. Elevated temperatures also promoted carbon loss through CO gas formation, while improved densification enhanced carbon-carbide contact, enabling further reaction. Hence, increasing SPS temperature and reducing carbon additions both drove cleaner microstructures without sacrificing densification.

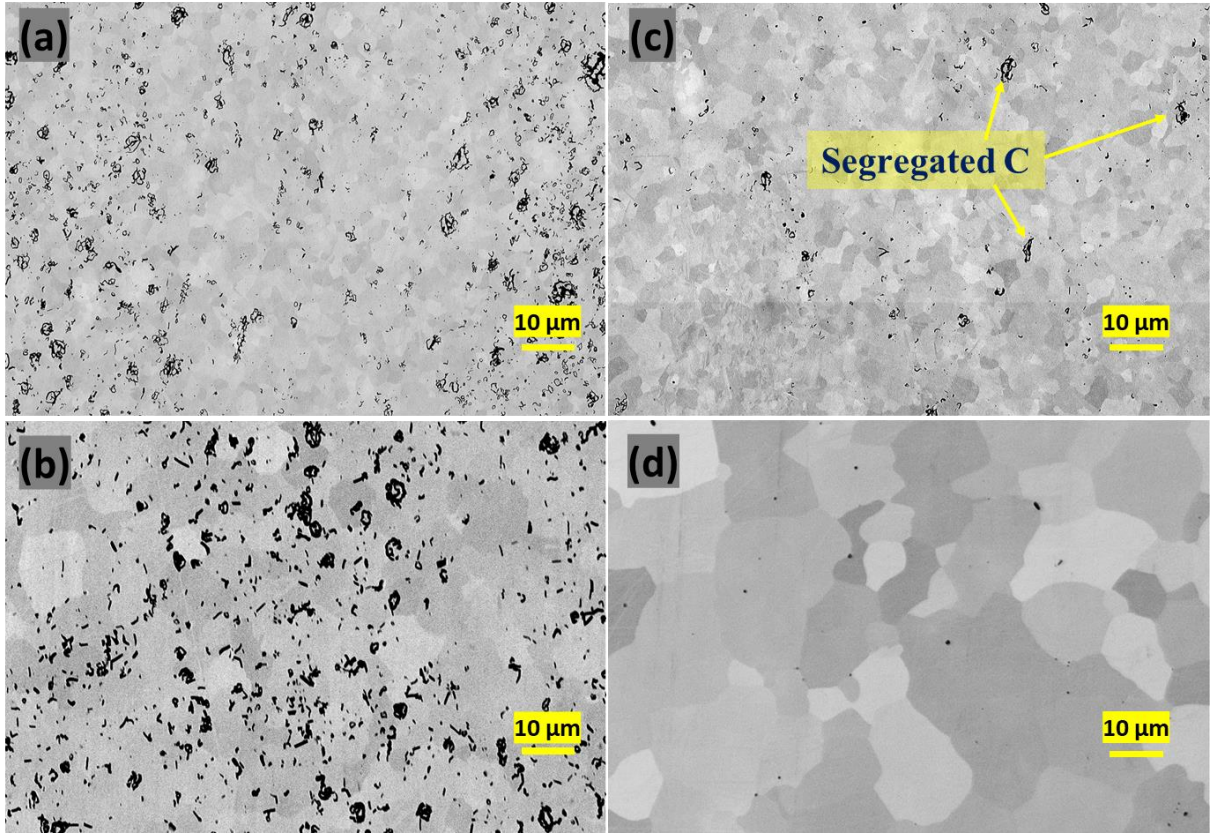

**Figure 2.** SEM micrographs of the (Cr,Mo,Ta,V,W)C synthesis optimization; (a) & (c) sintered at 1700 ºC, and (b) & (d) sintered at 1950 ºC with sub-stoichiometric batch carbon of 2.5 wt% for (a) & (b), and 7.5 wt% for (c) & (d).

The transition metals were distributed uniformly in the ceramics. Elemental maps for the optimized specimen, i.e., specimen 4, show spatially uniform signals for Cr, Mo, V, W, and C across large fields of view (Fig. 3). No oxide-rich or secondary-phase regions were evident; however, Ta segregation was observed for all specimens at length scales below the average grain size. Quantitative EDS performed over multiple regions yielded equiatomic metal contents of ~20 ± 3 at% for each of the five transition metals in every specimen, in close agreement with the batched ratio. Some differences in contents of the metals can be attributed to differences in detector efficiency, interaction volume, and minor local inhomogeneities. Similar uniformity was observed across all four compositions/temperatures, indicating that neither the increased carbon deficiency nor the higher SPS temperature drove measurable metal segregation. In sum, EDS mapping and quantification confirm equiatomic homogeneity and the absence of secondary metallic or oxide phases, strengthening the single-phase interpretation from XRD.

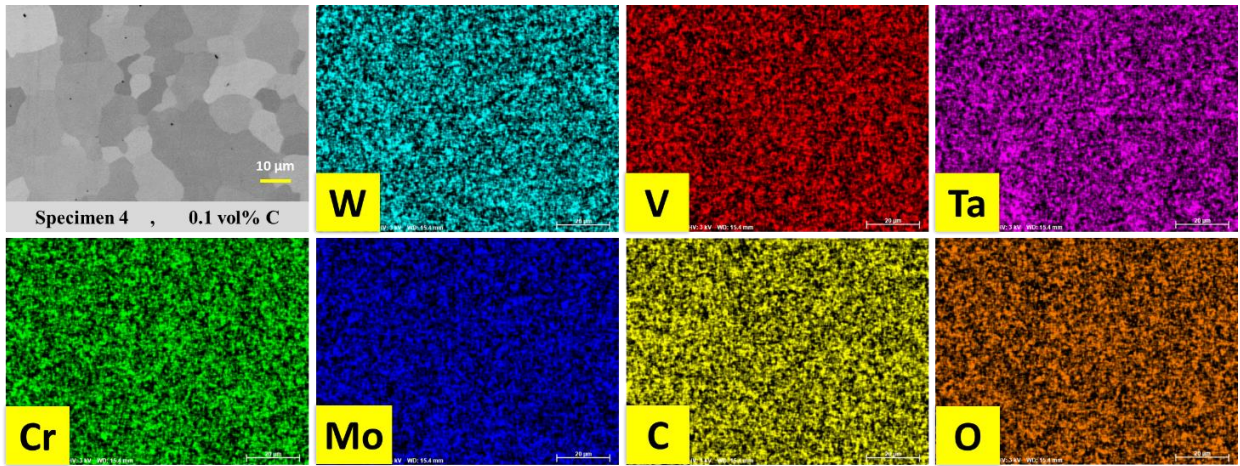

**Figure 3.** EDS elemental maps of the optimized (Cr,Mo,Ta,V,W)C specimen (7.5 wt% sub-stoichiometric batch carbon, sintered at 1950 °C).

*Thermal and Electrical Properties*

Thermal diffusivity increases approximately linearly from room temperature to 200 °C, (Fig. 4a). For the optimized microstructure, $D \approx 0.0285$ cm²/s at 200 °C, comparable to prior reports for related HECs as well as a high-entropy alloy with similar composition [31]. This positive slope reflects the faster growth of thermal conductivity with respect to heat capacity with temperature, which can be estimated from the graphs; thus, all compositions exhibit similar temperature dependencies. Heat capacity estimated by the Neumann-Kopp rule increased monotonically with temperature, and combining $C_p$, D, and density yields $\kappa \approx$ 7-9 W/m•K at room temperature rising to $\approx$ 10-12 W/m•K at 200 °C (Fig. 4b). Across compositions, decreasing microstructural excess carbon led to decreases in D and $\kappa$. The decreases were consistent with increased point-defect scattering when carbon vacancies were more prevalent, while boundary carbon can modestly assist heat flow in this temperature window by mitigating vacancy disorder. Therefore, within the temperature range of 25°·C to 200 °C, $\kappa$ primarily follows temperature and secondarily reflects a balance between vacancy and boundary scattering set by processing.

Room-temperature electrical resistivity decreased from ~137 µΩ•cm to ~120 µΩ•cm as excess carbon decreased from 5.4 vol% to 0.1 vol%, indicating reduced insulating boundary pathways and less electron scattering by amorphous carbon (Fig. 4c). Using the Wiedemann–Franz relation with the theoretical Lorenz number, the electron contribution to $\kappa$ was in the range of ~5.3-6.2 W/m•K, corresponding to ~65% (carbon-rich) up to ~88% (optimized) of the total thermal conductivity. Thus, curtailing boundary carbon improved charge transport and increased the electronic fraction of heat conduction at room temperature.

The trends in diffusivity and $\kappa$, together with changes in $\rho_e$, imply a competition between vacancy-driven scattering (from carbon-deficient lattices) and interfacial scattering (from amorphous carbon at boundaries). Higher SPS temperature and lower carbon contents reduced interfacial resistance, while excessive carbon deficiency can raise vacancy scattering; hence, tuning the amount of residual carbon in the microstructure and SPS temperature can be used to balance these mechanisms for optimal thermal transport.

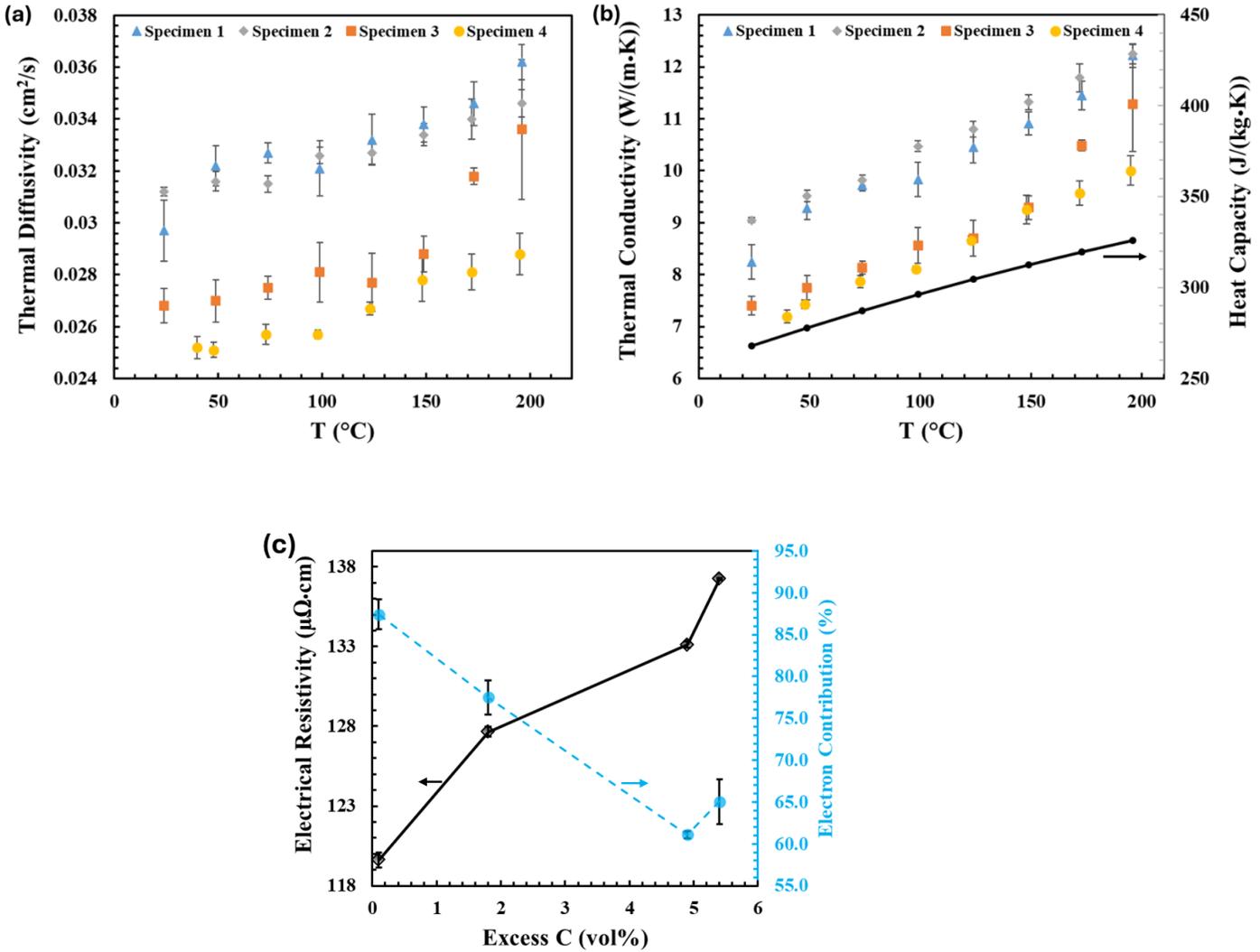

**Figure 4.** Thermal and electrical properties of (Cr,Mo,Ta,V,W)C HEC ceramics. (a) Thermal diffusivity. (b) Thermal conductivity and heat capacity (c) Room temperature electrical resistivity as a function of excess carbon in the microstructure (solid line) with the corresponding electron contribution to thermal conductivity (dashed line).

*Hardness*

All specimens had $HV_{0.05} \approx 28$-$29$ GPa at the lowest load (0.49 N), with larger uncertainties at low loads due to optical resolution and contrast thresholds (Fig. 5). No systematic dependence on batch carbon content, SPS temperature, or relative density were observed across the measured load range. The measured values fell within reported HEC hardness bands. Minor cracking was observed in some specimens along the diagonal directions for some indents, but most specimens did not exhibit significant cracking. Therefore, the (Cr,Mo,Ta,V,W)C system maintained robust hardness across the tested processing space, indicating mechanical performance was insensitive to the microstructural adjustments that affected thermal transport.

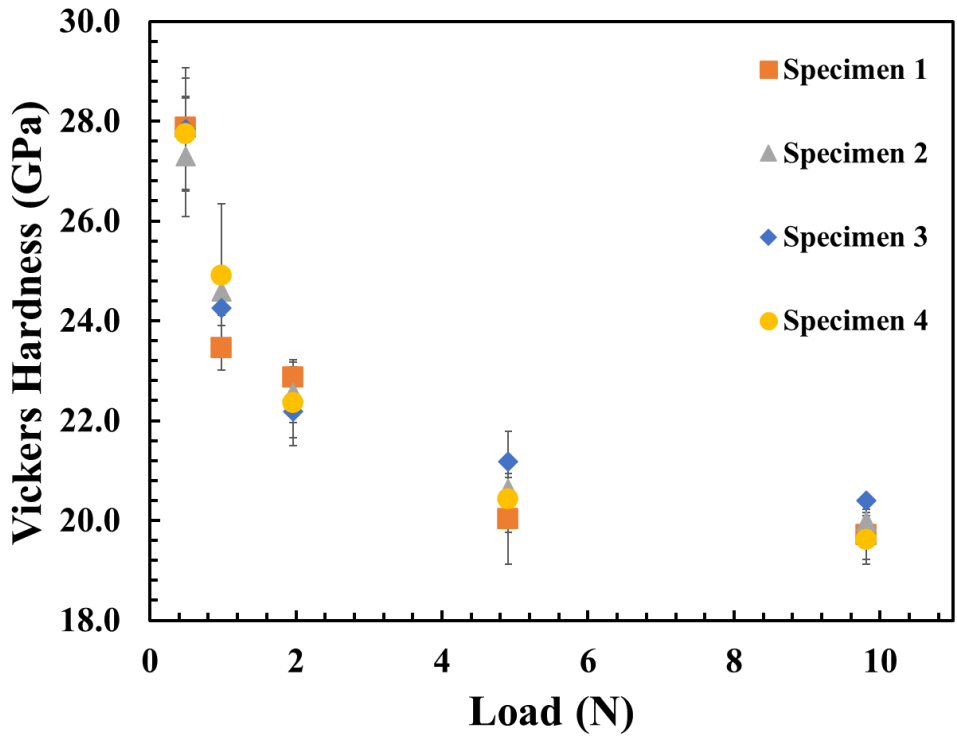

**Figure 5.** Vickers hardness as a function of indentation load for the (Cr,Mo,Ta,V,W)C specimens.

**Conclusions**

Single-phase, fully dense (Cr,Mo,Ta,V,W)C high-entropy carbide ceramics were produced by carbothermal reduction synthesis followed by spark plasma sintering. Two different densification temperatures were used, 1700 °C and 1950 °C, along with different amounts of carbon added to the initial batches. X-ray diffraction analysis confirmed that the ceramics had the face-centered cubic rock salt crystal structure, with lattice parameters increasing from 4.402 Å when the ceramic contained 5.4 vol% excess carbon to 4.409 Å when the ceramic contained 0.1 vol% excess carbon. Chemical analysis confirmed that the constituent metals were uniformly distributed in the microstructure and only carbon was observed as a secondary phase. Thermal diffusivity rose linearly with temperature, reaching ~0.0285 cm²/s at 200 °C. Calculated thermal conductivity increased from ~7-9 W/m•K at room temperature to ~10-12 W/m•K at 200 °C. Electrical resistivity decreased from ~137 μΩ•cm to ~120 μΩ•cm as carbon content in the microstructure decreased, leading to an increase in electron contribution to thermal conductivity from ~65% to ~88% of the total conductivity. All specimens exhibited a Vickers hardness of ~28-29 GPa at an indentation load of 0.49 N. These results demonstrated that controlling carbon content in the final ceramic and sintering conditions enabled tuning of the microstructure and transport properties in (Cr,Mo,Ta,V,W)C ceramics.

**Acknowledgements**

The Office of Naval Research MURI sponsored this research with the Grant Number of N00014-24-1-2768. The authors are grateful for Dr. Eric Bohannan at Missouri S&T for the helpful support and discussion with XRD.

**Contributions**

**A. Sarikhani** (Conceptualization, Methodology, Visualization, Validation, Writing – original draft, Data curation). **S. M. Smith** (Conceptualization, Methodology, Writing – review & editing). **S. Filipovic** (Conceptualization, Methodology), **W. G. Fahrenholtz** (Conceptualization, Supervision, Validation, Writing – review & editing), **G. E. Hilmas** (Supervision, Validation, Writing – review & editing).

principal transition metal carbides with high toughness and low thermal diffusivity. Appl. Phys. Lett. 114 (2019) 011905. https://doi.org/10.1063/1.5054954